\documentclass[review,a4]{elsarticle}

\usepackage[fleqn]{amsmath}
\usepackage{lineno,hyperref}
\usepackage[utf8]{inputenc}
\usepackage{lscape}
\usepackage{subfig}
\usepackage{longtable}
\usepackage{threeparttable}
\usepackage[font=scriptsize,labelfont=bf]{caption}
\usepackage{adjustbox}
\modulolinenumbers[5]
\usepackage{graphicx}
\begin{document}
\begin{frontmatter}
\baselineskip=.5cm
\title{Comparing the collective behavior of banking industry in emerging markets versus mature ones}
\author[a1]{H.~Vahabi\fnref{1}}
\author[a1]{A.~Namaki}
\author[a1]{R.~Raei}
\fntext[1]{Corresponding author.\\E-mail address:Hanie.Vahabi@ut.ac.ir}
\address[a1]{Department of Finance, Faculty of Management, University of Tehran, Tehran, Iran}
\begin{abstract}
\baselineskip=.5cm
One of the most important features of capital markets as an adaptive complex networks is their collective behavior. In this paper, we have analyzed the banking sectors of 4 world stock markets,which composed of emerging and matures ones. By applying one the important complexity notions, Random matrix theory(RMT), it is founded that mature markets have a higher degree of collective behavior,Even though we used RMT tools: participation ratio(PR), node participation ratio(NPR)and relative participation ratio(RPR) , which NPR illustrated independent banks than whole market and RPR compared collective behavior of markets by a normal range. By applying local and global perturbations, we concluded that mature markets are more vulnerable to perturbations due to the high level of collective behavior. Finally, by drawing the dendrograms and heat maps of the correlation matrices,we reaffirmed the stronger cross-correlation in the mature markets.
\end{abstract}
\begin{keyword}
\baselineskip=.5cm
\sep Cross-Correlation Matrix\sep Random Matrix Theory\sep Collective Behavior\sep Global Perturbation
\end{keyword}
\end{frontmatter}


\section{Introduction}
\baselineskip=.5cm
\indent Financial markets include components that have a lot of interaction. This interaction suggests that markets are examples of adaptive complex systems. A notable feature of these systems is the spatial and temporal dependence of their components\cite{johnson2003financial,stanley2000introduction,namaki2011network,albert2000error}. One of the most important characteristics of adaptive complex systems is their collective behavior. So in financial markets,(as an adaptive complex system) there is collective behavior. An individual study of a stock is not enough to predict its behavior. you need to consider the financial information of the entire market. This feature reflects the concept of collective behavior in markets\cite{mobarhan2016network}.There are several ways to look at collective behavior\cite{li2016determination,peron2011collective}. But a common approach to ecophysics to study collective behavior in markets is the analysis of cross-correlation matrix(C) of stock returns, which is analyzed by random matrix theory(RMT)\cite{mantegna1999introduction,potters2000theory,mehta2004random,guhr1998random}.The RMT was developed to explain the statistical properties of the energy levels of complex quantum systems in nuclear physics\cite{wigner1951statistical}. The basis of RMT is the study of the behavior of eigenvalues and their corresponding eigenvectors\cite{laloux1999noise,namaki2011comparing2}. This theory states that eigenvalues of the correlation matrix  are divided in two groups: noise and information. noises correspond to the eigenvalues of the random matrix, the area that they are located is called the bulk region and the information are largest eigenvalues, which are outside the bulk region \cite{plerou1999universal,plerou2001collective,saeedian2019emergence}. largest eigenvalue of the correlation matrix contains the market information and has the same effect on the whole system, which is called market-wide effect\cite{utsugi2004random,plerou1999scaling}.By eliminating the market-wide effect, the collective behavior of the market changes\cite{namaki2011network}.
 One of the methods based on RMT is  shuffling the off-diagonal elements of the correlation matrix C. This method eliminates the pattern of correlation between market components and thus the collective behavior of the market\cite{saeedian2019emergence,podobnik2010time}. Comparing the statistical characteristics of the C and shuffled C, we can provide valuable information about the collective behavior of the market, whatever the difference between the two matrices is less, the collective behavior of the market is weaker.\cite{saeedian2019emergence,jamali2015spectra,namaki2011comparing2}. Another  statistical tools of RMT is participation ratio(PR)\cite{bell1970atomic}. Participation ratio is a tool for estimating the number of significant participants in an eigenvector of a matrix\cite{pan2007collective}, we also use relative participation ratio(RPR) and node participation ratio(NPR)tools, RPR is used to measure the degree of collective behavior of each market and can be used to rank different markets in terms of collective behavior and NPR determines the share of each market component in the collective behavior of the market\cite{mobarhan2016network,saeedian2019emergence}.

\indent As mentioned above One of the features of complex systems is the spatiotemporal interdependence between the components of the system that change with the placement of one component to another. In the study of lim et al \cite{lim2009structure}, these events were applied to the correlation matrix in both local and global perturbations. They emphasized that markets are more sensitive to global perturbation and in another study \cite{namaki2011comparing}they expressed mature markets are more sensitive to global perturbation than emerging markets.

\indent The goal of this paper is to analyzed the degree of collective behavior on banks active in the Tehran stock exchange(TSE) and regional banks sector in SSE180 index as emerging markets and compare them with the regional banks sector active in Standard and Poor 500 index(S\&P500) and Nikkei225 index as mature markets. Then we measure the robustness of these banks against local and global perturbations, and whether is there a relationship between the degree of collective behavior of each market and its vulnerability to perturbation? The data under study included 21 regional banks(on average)active in the indices that were collected over a 3-years period from March 2016 to March 2019.

\indent This paper is organized as such after the introduction, the models used in the research are briefly presented then modeling results are shown and at the end of the paper conclusion have been done.
\section{Method}
\subsection{Cross-correlation matrix}
\baselineskip=.5cm
In order to calculate the correlation of a pair of stocks, we first need to calculate the logarithmic returns of two time series (two-stocks prices). Return $R_{i}(t)$ is defined as follows:
\begin{equation}
R_{i}(t)= log P_{i}(t+\Delta t)-log P_{i}(t)\label{Return}
\end{equation}
In the Eq.\eqref{Return}, $P_{i}(t)$ refer to the price of stock $i$ at the time of $t$. In this paper, we consider the $\Delta t$ is one day as the daily prices of each stocks are collected. To exclude the large effect of price on correlation coefficient, we use the normalized returns $r_{i}(t)$  defined as follows.
\begin{equation}
  r_{i}(t)= \frac{(R_{i}(t)-\langle R_{i}\rangle)}{\sigma_{i}}\label{Sreturn}
\end{equation}
In Eq.\eqref{Sreturn},$ \sigma_{i}$ is the standard deviation of the $R_{i}$, and $\langle \cdots \rangle$ denotes a time average over the period studied. The cross-correlation coefficient is defined as follows:
\begin{equation}\label{correlation coefficient}
  C_{ij}=\langle r_{i}(t)r_{j}(t) \rangle
\end{equation}
The amount of $C_{ij}$is in the range of $[-1,1]$ and the amount of $C_{ij}$ and $C_{ji}$ are equal\cite{namaki2011comparing}.
\subsubsection{Shuffled cross-correlation matrix }
Before defining Relative participation ratio(RPR), we must first introduce the shuffled cross-correlation matrix (shuffled C),the matrix C can be a diagonal matrix which means that there is no relationship between market components, but non-diagonal elements of the matrix C may be non-zero, indicating a correlation between market components, non-zero elements of matrix C are a necessary for collective behavior but not a sufficient condition\cite{saeedian2019emergence}. For the emergence of collective behavior in the market. In fact, we do not expect to behave collectively in a market where their constituents are in a completely random way. Therefore, in addition to having correlation among market components, some kind of pattern or structure for that correlation is essential. To create the shuffled C matrix we randomly displace the non-diagonal components of the matrix C this new matrix is called $C_{sh}$. By changing the matrix C any particular pattern of correlation disappears in off-diagonal regions of C but the relationships between individual elements of C are maintained. So the matrix $C_{sh}$ is a matrix where only the individual elements of the market components are correlated and the specific pattern of correlation in the off-diagonal regions is lost.\label{Csh}
\subsection{Relationship between RMT distribution and correlation matrix}
\baselineskip=.5cm
As mentioned above, the $C_{ij}$ is important for analyzing the relationship between a pair of stocks, However in this study we use the empirical approach. in previous work \cite{gopikrishnan1999scaling} it was shown that the correlation matrix C is separated into two parts. The part that corresponds to the RMT predictions and called noise and the other that deviates from the RMT called the information. To separate the noise and information parts The eigenvalues of the correlation matrix C are analyzed. It has been shown in previous work\cite{namaki2011comparing,gopikrishnan1999scaling}, that few of these eigenvalues are too far apart from others that these eigenvalues are the same information and their respective stocks have a wide effect on the whole market, which is called the market-wide effect\cite{laloux1999noise,lim2009structure}.
\subsection{Participation ratio}
\baselineskip=.5cm
Participation ratio's statistical tool was first introduced by \cite{bell1970atomic}. In the context of atomic physics, which later came into financial physics\cite{guhr1998random,plerou2002random} and is used to measure the degree of collective behavior in markets. In the diagonal matrix $C_{N\times N}$  gives us a set of eigenvectors $(u_{k})$ and eigenvalues $(\lambda_{k})$. eigenvalues show the collective mode of the market. Participation ratio (PR) for the $k_{th}$ stock is defined as follows:\label{PR define}
\begin{equation}
 P_{k}=(\displaystyle{\sum_{l=1}^N[u_{k}(l)]^4})^{-1} \label{PR}
\end{equation}
Where $u_{k}(l),l=1,\dots,N$ are the components of $u_{k}$. For the eigenvector even if it has a non-zero component, the value of PR from above is limited to N and from below is limited to unit. Then it can therefore be concluded that the amount of PR depends on N, (the size of the market under investigation). Therefore, it is necessary to eliminate the dependency PR of the market size and normalize it. For this reason, a new relative participation ratio(RPR) parameter is defined, so that the PR does not depend on the size of markets.\label{RPR define}
\subsubsection{Relative Participation Ratio}
\baselineskip=.5cm
The parameter Relative participation ratio RPR to eliminate dependency of PR is defined as the size of the matrix. The amount of $PR_{sh}$ is calculated according to the section\eqref{Csh} and \eqref{PR define}. Then amount of RPR defined as follows:
\begin{equation}
\delta=\frac{\langle PR_{sh} \rangle-\langle PR \rangle}{\langle PR_{sh} \rangle}\label{RPR}
\end{equation}
where $\langle PR \rangle$  and $\langle PR_{sh} \rangle$ represent average of PRs For all the eigenvectors of C and $C_{sh}$ respectively.
If the answer to Eq.\eqref{RPR} is close to zero it means that it is no different in C and $C_{sh}$. So it is implied that there is no specific pattern of correlation in the off-diagonal regions in  the matrix C so that the collective behavior of the market is also weak. Conversely, if the answer to the Eq.\eqref{RPR} is far from to zero, the specific pattern of correlation is strong and as a result, the collective behavior of the market is strong.
\subsection{Node Participation Ratio}
\baselineskip=.5cm
While the degree of collective behavior in the market is important, the contribution of each market component to this collective behavior is also questionable. So the parameter NPR is defined that as follows:
\begin{equation}
 N_{l}=(\displaystyle{\sum_{k=1}^n[u_{k}(l)]^4})^{-1} \label{NPR}
\end{equation}
$ N_{l}$ determines share of $l_{th}$ component in the collective behavior of the market. Since PR examines the contribution of a eigenvector $u_{k}$ on the basis of a eigenvalue $\lambda_{k}$ in the collective behavior of market. While NPR examines contribution of $l_{th}$ market component in it. NPR reflects the independence of the companies than market. This means that a company with a lower NPR has a higher independence than a company with a higher NPR. $N_{l}^{-1}$ measures the independence of the company $l_{th}$ from the rest of the companies\cite{saeedian2019emergence}.
\subsection{Perturbing a correlation matrix}
\baselineskip=.5cm
\indent As summarized above, applying the perturbation to the correlation matrix eliminates some of the stylized features observed in financial time series, such as the genuine correlation between two stocks belonging to the same business group\cite{lim2009structure}. In the following this section, we describe the details of the perturbation on the correlation matrix, Firstly, the local perturbation:
\begin{enumerate}
\item select randomly one of the off-diagonal components of the cross-correlation matrix.
\item generate two Gaussian-distributed time series (white noise series) and calculate their cross-correlations.
\item substitute the calculated correlation in step 2 instead of the original correlation selected in step 1.
\end{enumerate}
Because of the symmetry of the cross-correlation matrices, $C_{ij}$ and $C_{ji}$ are equal.
So, when local perturbation is applied, they are both placed simultaneously. As mentioned, local perturbation is a technical study\cite{lim2009structure}. It has no practical value.The meaning of global perturbation is its overall impact on market returns, so the global perturbation has a stronger and pervasive effect. This perturbation applies as follows:
\begin{enumerate}
\item select a off-diagonal components of the cross-correlation matrix.
\item Identify the two stocks  belonging to the correlation coefficient of Step 1.
\item generate two Gaussian-distributed time series.
\item Instead of the two stocks identified in step 2, place two Gaussian-distributed time series in the return matrix.
\item From the modified return matrix, calculate the new cross-correlation matrix.
\end{enumerate}
Steps 4 and 5 create a global perturbation that is related to the pair of stocks selected in Step 1
to add a more structured account of the behavior of stocks to global perturbation, we add a rule to the step 1 instead of randomly selecting the correlation coefficient, we first choose the pair of stocks with the strongest correlation rather than the weakest correlation. This is called top-ranked method, and vice versa, from the weakest to the strongest correlation, this is called bottom-ranked method.\label{GL introduce}
\section{Results}
\begin{figure}
  \centering
  \includegraphics[width=1\textwidth]{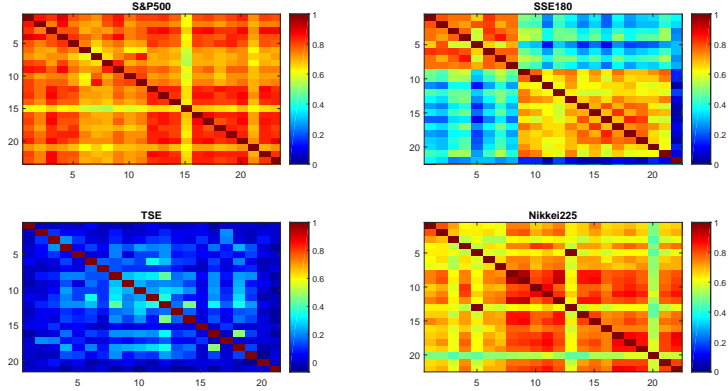}
  \caption{Color image of correlation matrices.Warmer colors show higher cross-correlation between components.}\label{colorimage}
\end{figure}

\begin{figure}
\centering
\baselineskip=.1cm
\includegraphics[width=1\textwidth]{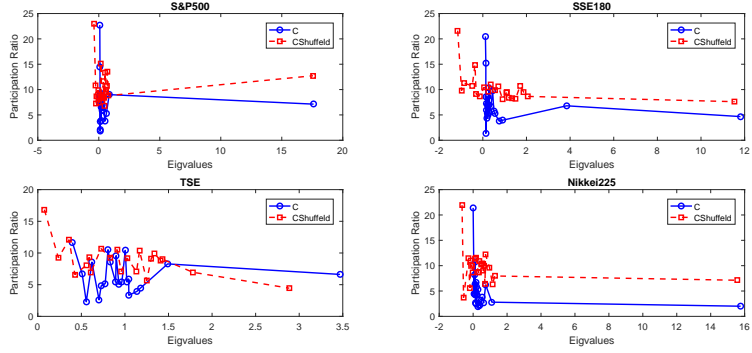}
\caption{Compare PR and $PR_{sh}$ for active banks in  indices. Although PR and $PR_{sh}$ in S\&P500 and Nikkei225 behave more similarly, the collective behavior of the these markets is stronger.}\label{fig1}
\end{figure}

\begin{figure}
\centering
\includegraphics[scale=.35]{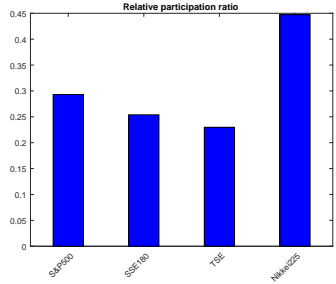}
\caption{RPR is shown for  indices and it is higher for the mature markets.}\label{fig2}
\end{figure}

\baselineskip=.5cm
The purpose of the methods presented in this study is to compare the collective behavior of banks in the 4 indices S\&p500,SSE180,Nikkei225 and TSE. In these 4 indices, SSE180 and TSE  are active in emerging markets and S\&P500 and Nikkei225 are in the mature markets. We will also analyzed the behavior of these banks to the perturbation. For introduce, we see in fig.\eqref{colorimage} color images of correlation matrix C for 4 markets 2 mature markets have higher correlation. That SSE180 is stronger than TSE, and it is moderate,however TSE as weak market has lower correlation. In fig.\eqref{fig1}, compare the PR and $PR_{sh}$ for the 4 indices. If PR and $PR_{sh}$ behave closely together, the collective mode is weak,SSE180 and TSE have closely behavior then matures, so we can consider emerging markets in collective mode are weak. in fig. \eqref{fig2} the $\delta$ value is shown, as stated, if the $\delta$ value is near to zero The collective behavior of the market is weak and it is far from zero, it is strong. So it can be concluded that the collective behavior of each of the 4 indices is weak, but it is stronger in the two mature indices,as shown in the paper \cite{saeedian2019emergence}.

\begin{figure} 
    \centering
    \includegraphics[scale=0.35]{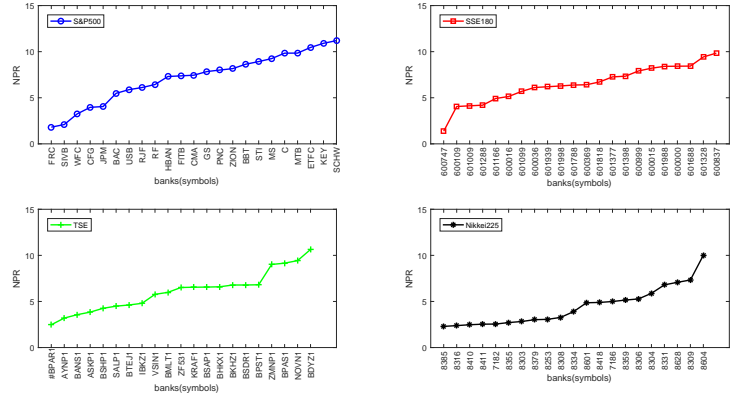}
    \caption{NPR for whole banks in Indices. Banks with a higher NPR have a higher contribution in the collective behavior of the market, and vice versa.}\label{fig4}
\end{figure}

\begin{figure}
  \centering
  \includegraphics[scale=0.35]{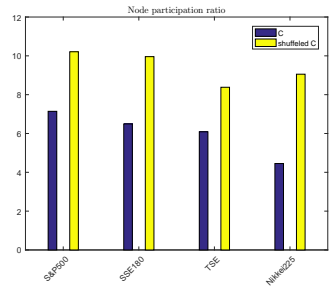}
  \caption{Comparison of NPR and $NPR_{sh}$ for the indices.}\label{fig5}
\end{figure}

 Then we show the collective behavior of these 4 indices by using Eq.\eqref{NPR}, we measure the contribution of each bank to the collective behavior of the same index. In fig.\eqref{fig4} banks with lower NPR are more independent than the general trend of market and conversely, banks with higher NPR depend on other banks. In fig.\eqref{fig5} compares NPR and $NPR_{sh}$ in 4 Indices. Nikkei225 has less NPR. But its low NPR indicate this market behaves independently of other markets, and it is more immune to the dangers of the global shocks. It also in $C_{sh}$, NPR increased. So it can be concluded that the existence of a strong collective mode of that market.
\qquad

\begin{figure}
   \centering
\includegraphics[width=1\textwidth]{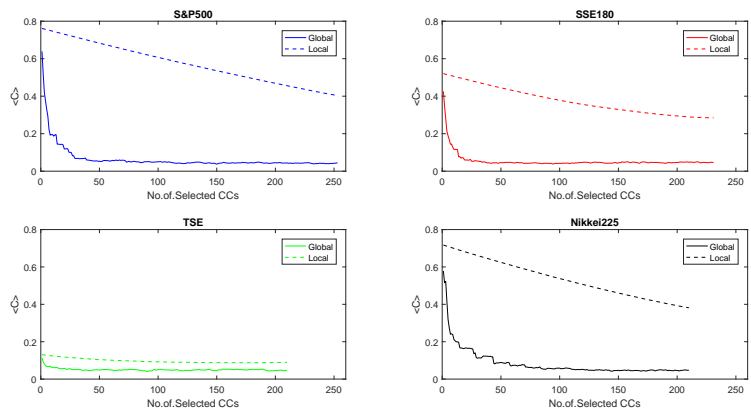}
\caption{Compare of global perturbation and local perturbation for indices.}\label{fig10}
\end{figure}

\begin{figure}
   \centering
\includegraphics[width=1\textwidth]{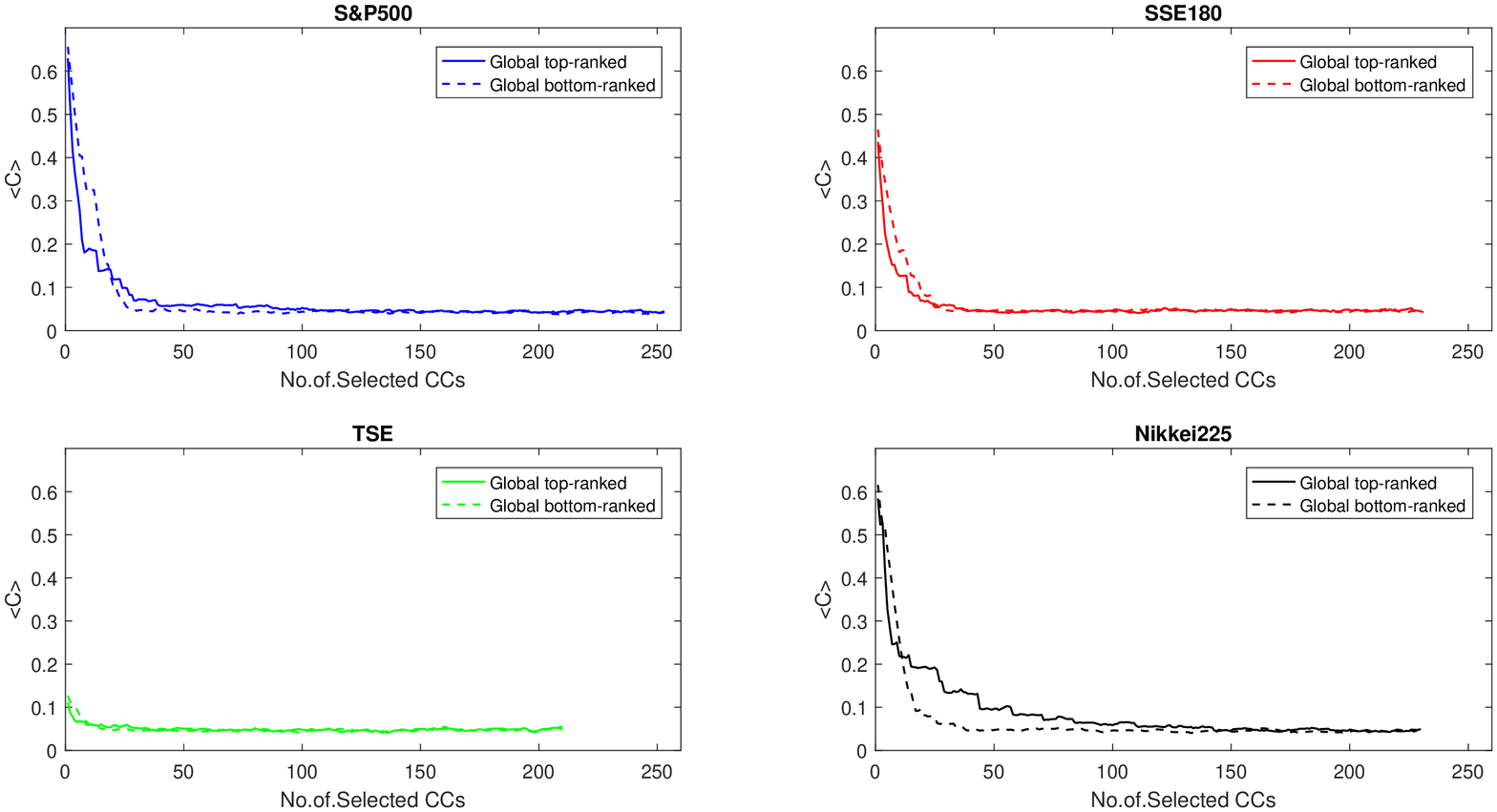}
\caption{Compare of global perturbation by two method top-ranked and bottom-ranked.}\label{fig7}
\end{figure}

\indent Another methods used in this study are local and global perturbation. In the fig.\eqref{fig10} global and local perturbations apply to all  indices. As stated in the paper \cite{namaki2011comparing} mature markets react more to this perturbations. Because the range of mean value of correlation coefficients changes is higher for mature markets. It can be said that the effect of global and local perturbations on mature markets is higher than on emerging markets. the TSE index has been less responsive to the perturbation as it is more isolated than the another markets, and it is almost internally active. Thus, markets with stronger collective behavior are more vulnerable to local and global perturbations. As mentioned in the section\eqref{GL introduce}there are two methods top-ranked and bottom-ranked to make the perturbation. In the fig.\eqref{fig7}the two methods are compared which show that they are not significantly different from each other.
\qquad

\begin{figure}
   \centering
\includegraphics[width=1\textwidth]{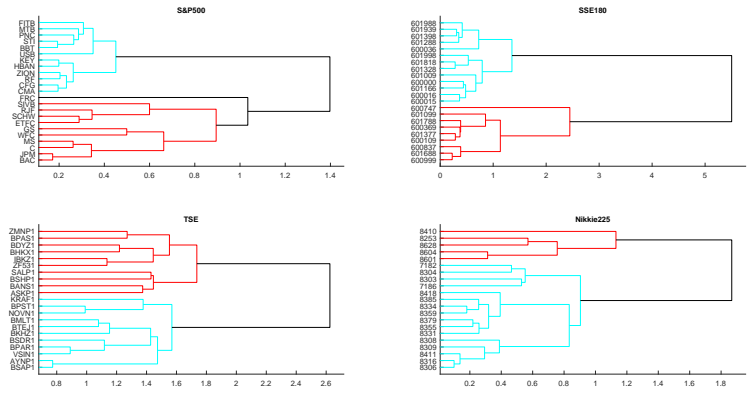}
\caption{Compare dendrogram of indices.vertical axis represents proximity of components. In mature markets, this criteria shows strong correlation and in TSE, it shows weak correlation but in SSE180, market is separated 2 sections. Each section has strong correlation but with another section is weak.}\label{fig8}
\end{figure}

\begin{figure} 
    \centering
    \subfloat[S\&p500 ]{\includegraphics[width=.35\textwidth]{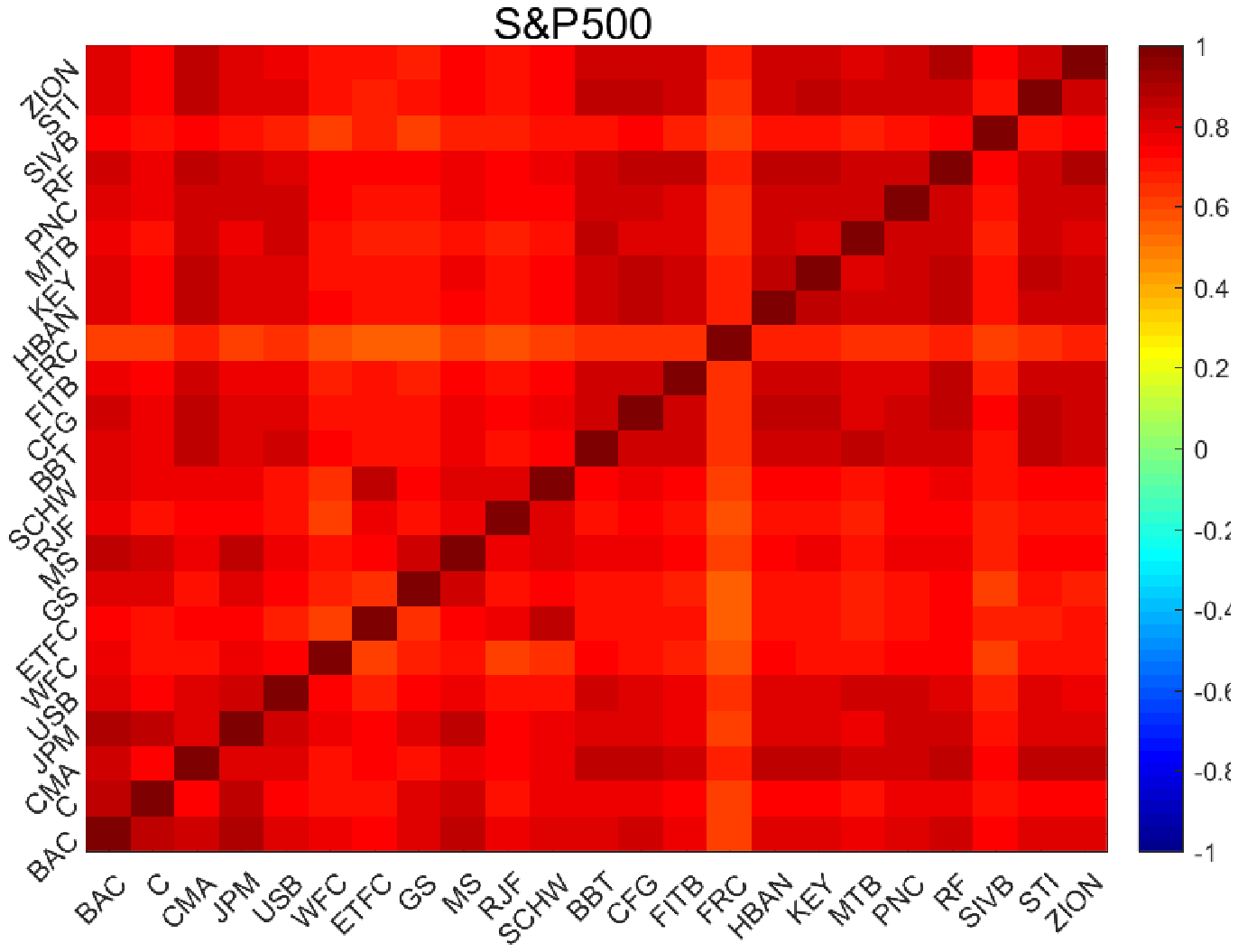}}%
    \qquad
    \subfloat[SSE180 ]{\includegraphics[width=.35\textwidth]{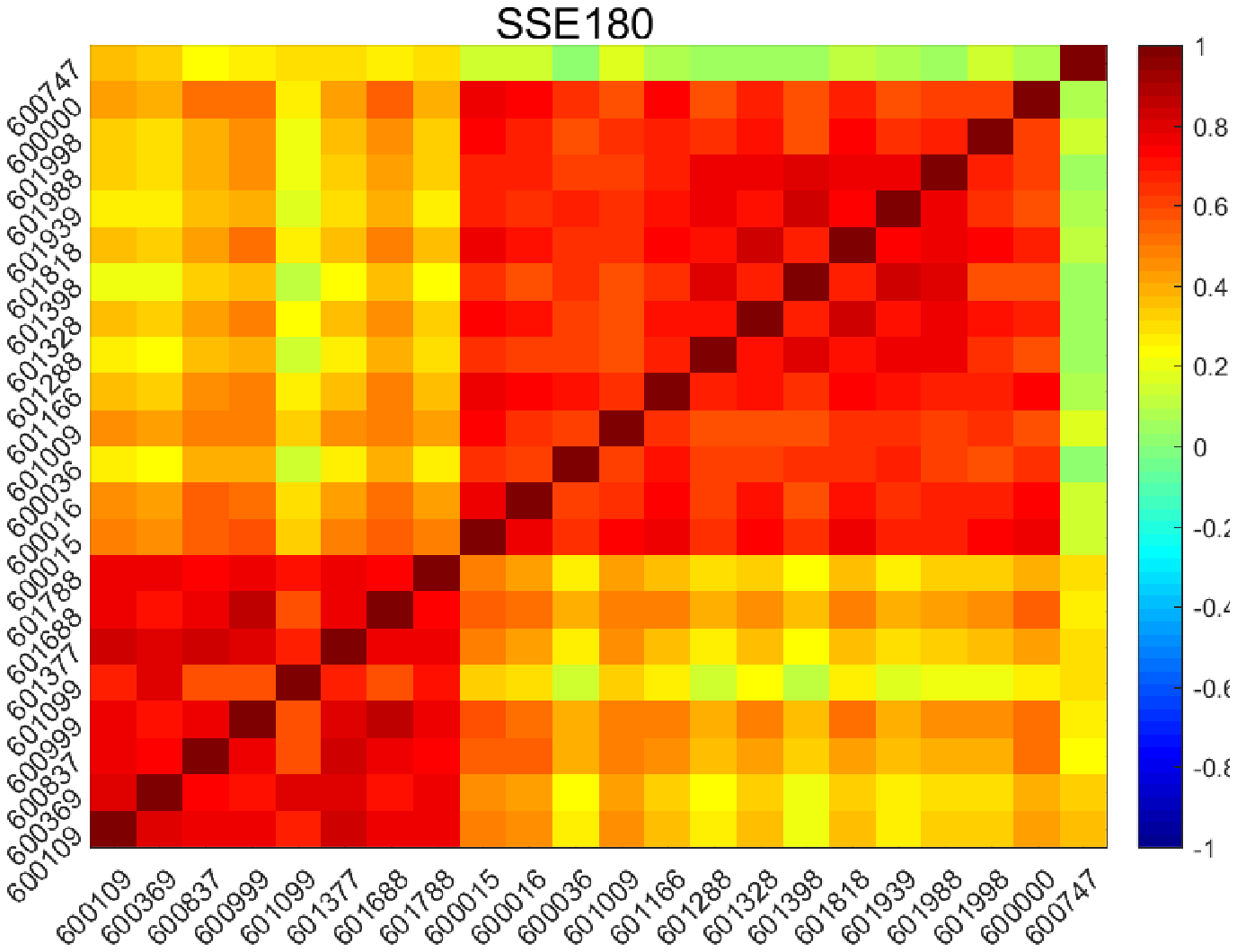}}%

    \subfloat[TSE ]{\includegraphics[width=.35\textwidth]{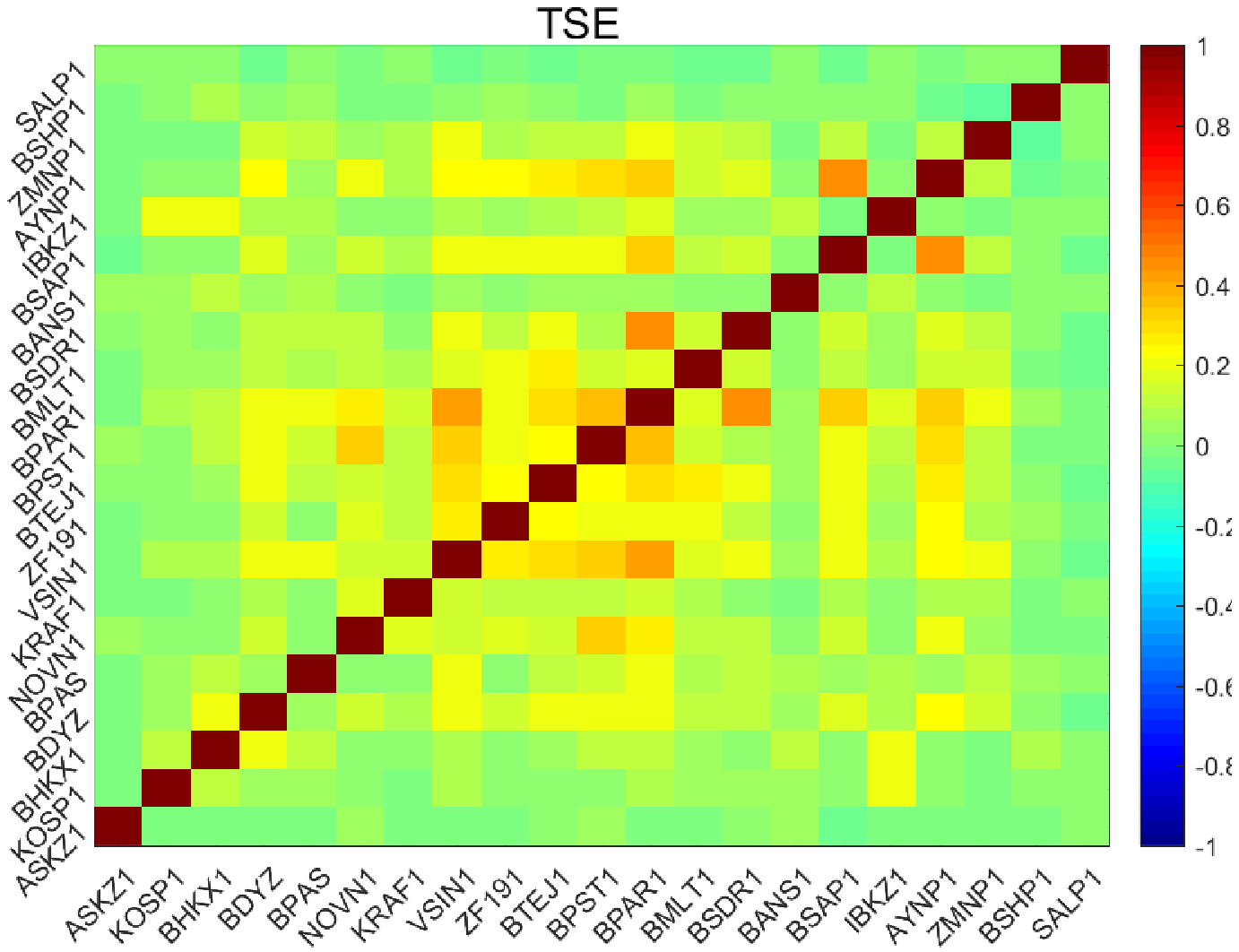}}%
    \qquad
    \subfloat[Nikkei225]{\includegraphics[width=.35\textwidth]{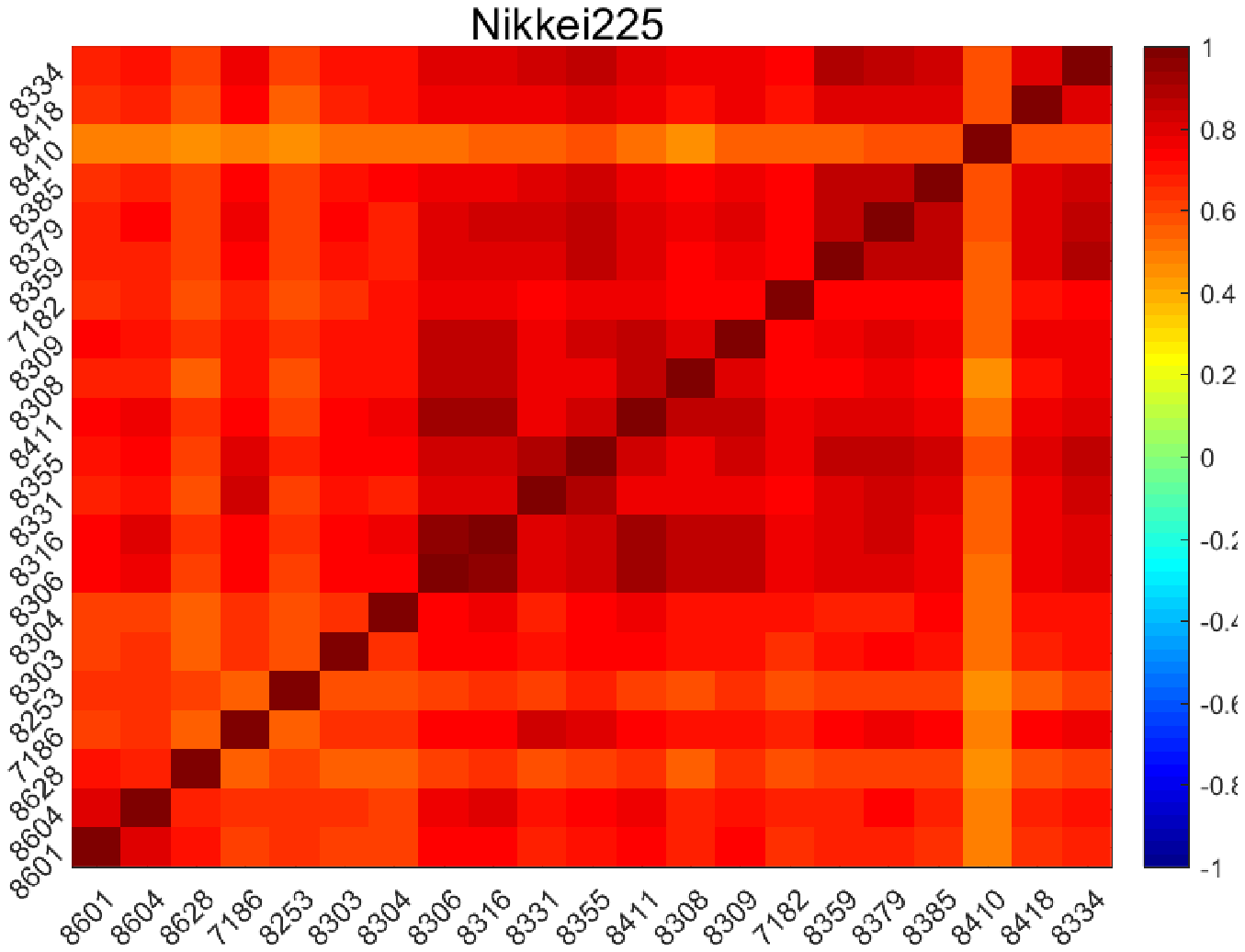}}%
    \caption{ Heat maps of indices.Heat maps with warmer colors have stronger correlation.}%
    \label{fig9}%
\end{figure}

\indent At the end of the discussion, we draw dendrograms and heat maps of cross-correlation matrices, . Both are derived from the clustering algorithm\cite{cattell1943description},and they are clustered in Ward's method\cite{ward1963hierarchical}. in dendrogram fig.\eqref{fig8} horizontal axis represents Proximity criteria of the correlation coefficients and the vertical axis represents the criterion of non-proximity of correlation coefficients As expected, mature markets have clustered more rapidly,which is concluded that cross-correlation coefficients of these markets is close, in emerging market, TSE clustered slower than matures. While the SSE180 is split into 2 completely separate clusters, which is illustrated in both the color image of the correlation matrix C and the heat map too. In heat maps, fig.\eqref{fig9} warmer colors represent stronger correlation,which can be seen that the matures have stronger correlation as seen in dendrograms.

\section{Conclusion}
\baselineskip=.5cm
In this study, we investigated two themes: the collective behavior of the banking sector in 4 indices and the extent of their robustness to global and local perturbations. As expected,degree of collective behavior of mature markets were stronger than emergings. The SSE180 index is active in a market with high growth rates that will soon become a mature market. Its collective behavior is not as strong as mature markets but not as weak as emerging markets. Therefore, this market behaves moderately. While Iran's index, TSE is a fully emerging market and the low correlation among its members proves it, RPR's value also indicates that is very weak in collective behavior.So the behavior of each stock can be predicted without considering the rest of the market in Iran's.

\indent As the collective behavior of mature markets is stronger than emerging markets, they are also more vulnerable to perturbations, so the collective behavior of the market is directly related to its vulnerability to perturbations. Furthermore the effect of local perturbation on all markets is weaker than global perturbation. The methods of perturbation on markets, Top-Ranked and Bottom- Ranked are the same. For future studies can analyze the correlation of world currencies and compare their collective behavior with their respective markets. We can also measure their sensitivity by applying perturbations, and do the currencies of developing countries have a weaker degree of collective behavior than the currencies of developed countries?

\section{Acknowledgments}

\section{Refrences}
\baselineskip=.5cm
\bibliography{Ref}
\end{document}